\documentclass[letterpaper]{JHEP3}
\usepackage{psfrag,amsmath,amsthm,amssymb,subfigure,array}
\usepackage{multirow}  
\usepackage{url}
\usepackage{ae}  
\usepackage[all]{xy}  
\usepackage{wrapfig}  

\ifpdf     
    \usepackage[pdftex]{graphicx}
    \usepackage{epstopdf}  
\else  
    \usepackage[dvips]{graphicx}
\fi

\newcommand{\Rb}{\overline{R}}
\newcommand{\gb}{\overline{g}}
\newcommand{\nb}{\overline{\nabla}}
\newcommand{\Fb}{\overline{F}}
\newcommand{\f}[1]{\frac{1}{#1}}
\newcommand{\be}{\begin{equation}}
\newcommand{\ee}{\end{equation}}
\newcommand{\bea}{\begin{eqnarray}}
\newcommand{\eea}{\end{eqnarray}}

\preprint{FERMILAB-PUB-09-598-T\\CERN-PH-TH/2009-236\\arXiv:0911.5343}
\title{Inflation on the Brane with Vanishing Gravity}
\author{Jason Gallicchio\\
Jefferson Physical Laboratory, Harvard University, Cambridge MA 02138\\
E-mail: \email{jason@physics.harvard.edu}}
\author{Rakhi Mahbubani\\
Theory Division, CERN, CH-1211 Geneva 23, Switzerland and \\
\;Fermi National Accelerator Laboratory, Batavia IL 60510\\
E-mail: \email{rakhi@mail.cern.ch}}
\date{\today}

\abstract{Many existing models of brane inflation suffer
from a steep irreducible gravitational potential between the
branes that causes inflation to end too early.  Inspired by the fact
that point masses in 2+1 D exert no
gravitational force, we propose a novel unwarped and
non-supersymmetric
setup for inflation, consisting of 3-branes in two extra dimensions compactified on a sphere.  The
size of the sphere is stabilized by a combination of a bulk
cosmological constant and a magnetic flux.  Computing the 4D
effective potential between probe branes in this
background, we find a non-zero
contribution only from exchange of level-1 KK
modes of the graviton and radion.  Identifying antipodal points on the
2-sphere projects out these modes, eliminating entirely the
troublesome gravitational contribution to the inflationary
potential.}

\begin{document}

\section{Introduction}

As a solution to the horizon, flatness, heavy relic and
structure formation problems in the early universe, the
inflationary paradigm \cite{Guth:1980zm}, especially in its
slow-roll incarnation \cite{Linde:1981mu,Albrecht:1982wi}, is
well-established and supported by recent observations
\cite{Leach:2003us}. However to date there is arguably no simple
and compelling canonical model in which slow-roll inflation arises
generically, although there is some hope that a viable
framework might be found in string theory (for reviews see
\cite{Quevedo:2002xw}).

A promising example of such a framework is brane inflation
\cite{Dvali:1998pa,HenryTye:2006uv}, in which the ostensibly
flat brane-brane potential, in extra dimensions large compared
with the brane thickness, was used for inflationary slow-roll.
Large extra dimensions, with all fields except gravity confined
to a brane, were originally proposed to explain the hierarchy
between the weak and gravitational scales
\cite{ArkaniHamed:1998rs}. In an inflationary context they
allow for sufficient separation between the branes to render
gravitational effects negligibly small, with the brane tensions
providing a large and constant contribution to the inter-brane
potential.  As the branes approach each other, the weak
Newtonian gravitational attraction between them is responsible
for an inflation-ending collision that could reheat the
universe.

This naive expectation was called into question by Kachru et
al., who argued in \cite{Kachru:2003sx} that finite volume
effects and moduli stabilization typically spoil the `flat'
gravitational brane potential.  They concluded that viable models of brane
inflation could not arise from {\it generic} string
constructions with all moduli stabilized, since a correct
treatment of the moduli generally renders brane potentials too
steep for inflation.

The idiosyncratic nature of gravity in 2+1 dimensions has been
known and understood since the 1960s: there is no
Newtonian gravitational potential due to a point mass in 2+1
dimensions \cite{Staruszkiewicz;1963}. Instead, the mass
cuts out a
deficit angle from the space in which it lives, giving rise to a
conical
surface that is flat everywhere apart from at its location.  A second
point mass placed anywhere on this cone feels no gravitational
force.  Similarly the gravitational potential between any two
codimension-2 objects in infinite flat
space: for example cosmic strings in 4 dimensions,
or 3-branes in 6 dimensions, is exactly constant.

Historically there has
been considerable phenomenological interest in two extra spatial dimensions.
In order to explain the gauge hierarchy, these naturally select
millimeter-sized dimensions, which just evaded all
limits from the best experimental tests of gravity at the time\footnote{Current constraints from astrophysics
 restrict the size of two extra dimensions to less than 0.16 nm
 \cite{Hannestad:2003yd, Amsler:2008zzb}.}.  The same scale also
corresponded with that of the cosmological constant and neutrino masses,
a coincidence that led to much model-building excitement.
For example, a 2-sphere compactification of the extra dimensions,
stabilized by a bulk cosmological constant and
magnetic flux was used by \cite{Sundrum:1998ns,Carroll:2003db,Chen:2000at} in order to explore the
cosmological constant problem.
We employ an identical configuration as a background in which to
explore brane inflation \cite{Lee:2009gn}, seeking a general unwarped mechanism to
flatten the steep irreducible gravitational contribution to
the inter-brane potential.  Unlike previous analyses which contain exact
solutions in the presence of finite-tension branes in specific
configurations,
we treat the branes as purely gravitational perturbations at arbitrary positions on the
stabilized 2-sphere.  We then compute the
solution to leading order in this perturbation by integrating out all
interacting massive KK modes in the 4D effective theory.

We begin this paper with a careful analysis of Kachru
et al.'s assertion that extra dimensional brane inflation does not
generically work (see Section \ref{sec:problems}).  We couch their
argument in terms of a shape function for the
inflaton potential, which makes it evident that the inflationary
parameter $\eta$
is independent of all dimensionful parameters.  We then explain,
from the perspective of the 4D effective theory, the need for a
volume subtraction in closed spaces.  This constrains $\eta$
to be an $\mathcal{O}(1)$ quantity, which is too large for successful
slow-roll inflation.  In Section \ref{sec:nonperturbativeflat} we
attempt to understand perturbatively the behavior of point masses in
2+1 D.  We subsequently examine two 3-branes in a flat 6D space in
Section \ref{sec:perturbativecodimension2}, and
find that the potential due to graviton exchange is exactly zero as
expected.  From the standpoint of the 4D effective theory, we find
that this lack of gravitational
potential results from a direct cancellation between
the potential due to the exchange of massive graviton and radion
modes.  We show in Section \ref{sec:FR} that this is no longer the
case in a
spherical background stabilized by a bulk cosmological constant and
magnetic flux.  Rather, the potential only vanishes for
$\ell\ge 2$ spherical harmonic modes; exchange of the $\ell=1$ modes gives
rise to a repulsive cosine potential that is again too steep,
ending slow-roll inflation early. This striking outcome suggests
a simple solution: identifying antipodal
points on the 2-sphere projects out odd KK modes, again
giving exactly zero potential between the branes.  We discuss two
possibilities for adding an interaction that would make this setup
the basis for a viable model of inflation: a massive bulk scalar coupling
to the branes, and the effect of the gravitational casimir between branes.
Finally we discuss the
effect of dialing up the brane tensions so they can no longer be
treated as perturbations on the spherical background, but instead
cut out finite deficit angles from the extra dimensional space.  We
predict zero gravitational potential even in this limit, and make some
observations about existing models with codimension-2 branes.

Throughout this paper we use the field theorist's mostly minus
metric sign convention $(+-----)$, with uppercase Roman
letters labelling 6D coordinates $X^M$ with metric $g_{MN}$;
lowercase Greek for extended 4D Minkowski coordinates
$x^\mu$ with metric $g_{\mu\nu}$; and lowercase Roman for
coordinates in the extra dimensions $y^m$ with metric $g_{mn}$.


\section{Pitfalls of Brane Inflation}
\label{sec:problems}

The original brane inflation scheme \cite{Dvali:1998pa} treated
branes as point particles in infinite flat extra dimensions.
Naively the potential due to two branes, a distance $r$ apart, can be thought of as
the sum of their brane tensions ($f^4$) and the Newtonian potential
between them in $n$ extra dimensions:
\be\label{equ:Newtonianpotential}
V(r) = f_1^4 + f_2^4 + \frac{1}{M^{D-2}} \frac{f_1^4 f_2^4}{r^{n-2}}
\ee
where $D$ is the total number of dimensions and $M$ is the
fundamental Planck scale. This potential looks remarkably flat
when the branes are far from each other since the tension terms
dominate over the negligible gravitational potential (see Fig.\
\ref{fig:braneinf}).

\FIGURE{
\psfrag{Vr}{$V(r)$}
\psfrag{BraneSize}{$\frac{1}{f}$}
\psfrag{r}{$r$}
\includegraphics[width=4.5in]{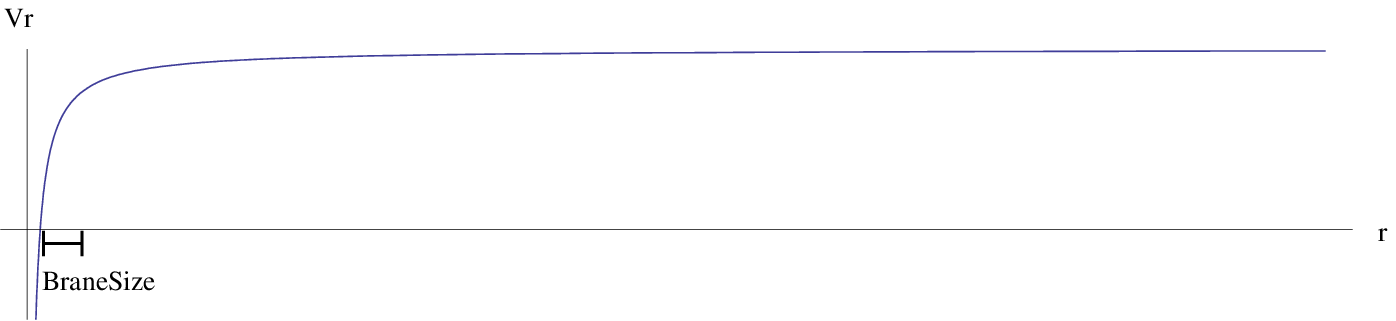}
\caption{The Newtonian potential between branes in infinite
flat extra dimensions is nearly constant when the branes are
far apart by comparison with their size ($1/f$), since the
gravitational attraction between them is negligible.}
\label{fig:braneinf}
}

For a large enough extra dimension compared
with the size of the branes ($1/f$), it seems like it should be
possible for the branes to get far enough away from each other
to sample the flat region of the potential.  However there are
two problems with this naive argument.  First, the slow roll
parameter $\eta$ also involves the 4D Planck mass; we will show
below that this results in a cancellation between all the
dimensionful parameters in the theory, leaving $\eta$ dependent
on a single dimensionless function of the angular separation of
the branes. Because of this, hierarchies between dimensionful
parameters cannot affect $\eta$.  Moreover, the Newtonian
potential in Eq.\ (\ref{equ:Newtonianpotential}) needs to be
modified for a compact space; we show why this is, and
how it prevents $\eta$ from being small enough for a successful theory
of slow-roll inflation.

Factoring out dimensionful quantities from the
potential between two branes of equal tension $f^4$:
\be
V(\theta) = 2f^4 + \frac{1}{M^{D-2}} \ \frac{f^8}{L^{n-2}} \ \varphi(\theta)
\ee
where $L$ is the size of the space and $\varphi(\theta)$ is a
dimensionless function describing the shape of the potential.

As argued in \cite{Sundrum:1998sj}, the Lagrangian for two
branes labeled by $i$ at positions $\vec Y_i$, is given by the
measure on each brane,
\be
\mathcal{L}
\ = \ \sum f^4_i\sqrt{g_{\mu\nu}(\vec Y_i)}
\ \simeq \ f^4 \,
\left(\partial_\mu \Delta \vec Y\right)
\cdot
\left(\partial^\mu \Delta \vec Y\right)
\ee
then the inflaton field $\Phi$ is the canonically normalized distance
between the branes, $|\Delta \vec Y|$
\be
\label{eqn:normalized_inflaton}
\Phi^2 = f^4 \, \Delta \vec Y^2 = f^4 \, L^2 \, \theta^2
\ee
The inflationary parameter $\eta$ is defined in terms of the
Hubble constant $H$ as follows:
\bea
\eta &\equiv& \frac{m_\Phi^2}{H}
     = M_4^2 \frac{V''(\Phi)}{V(\Phi)}  \nonumber\\
     &\simeq& \varphi''(\theta) \qquad \textrm{for large brane separation}
\eea
where $M_4$ is the 4D Planck mass, $M_4^2=M^{D-2}L^n$.  As
claimed, $\eta$ is independent of all dimensionful quantities,
and depends only on the shape of a dimensionless potential.

The Newtonian potential at position $\vec y$ in $n$ extra
dimensions, due to a brane with tension $f^4$ at the origin, is
governed by the Poisson equation:
\be
\label{equ:newtonian_potential_eqn}
\nabla^2 \phi(\vec y) = \frac{f^4}{M^{D-2}} \delta^{(n)}(\vec y)
\ee
This equation is inconsistent for a closed space since the left
side is a total derivative and so integrates to zero.
Kachru et al.\ modify this by subtracting a volume-dependent term
that ``emerges naturally from the curvature of the four-dimensional
space-time'' \cite{Kachru:2003sx}.
\be
\label{equ:potential_modified}
\nabla^2 \phi(\vec y) =  \frac{f^4}{M^{D-2}} \left( \delta^{(n)}(\vec y ) - \frac{1}{L^n} \right)
\ee
We argue that this correction actually emerges from a careful
treatment of the zero mode in the 4D effective theory, even in a
flat, closed space.

Recall that the low-energy effective potential between
two branes can be computed by integrating out all the heavy KK
modes coupling to them.  Zero modes coupling to the branes
however, should be treated differently: they must remain in the
effective theory, and will in general give rise to a
time-dependent cosmological expansion.  Hence the Green's
function for the potential in a compact space must include a
sum over all modes \emph{except} for the zero mode.  We
illustrate this point below using the example of the potential due to
point masses on an
$n$-torus\footnote{This is exactly equivalent
to finding the Newtonian
\emph{potential density} due to a codimension-$n$ brane.}, and show
that the seemingly ad hoc volume subtraction
corresponds exactly to leaving out the zero mode from the sum
over modes in the Green's function.

To solve for the potential due to a point mass $m$ on an
$n$-torus of radius $L$,  we
first transform the naive flat-space Poisson equation, Eq.\
(\ref{equ:newtonian_potential_eqn}), into momentum space to
obtain
\be
\vec{k}^2 \tilde{\phi}_{\vec{n}} = \frac{m}{M^{n+2}} e^{-i \vec{k}_n
  .\vec{y} }\qquad\qquad{\rm with}\qquad\qquad \vec{k}_n =
\frac{2\pi\vec{n}}{L}
\ee
for each fourier mode $\tilde{\phi}_{\vec{n}}$.
  This equation is
obviously inconsistent for $\vec{n}=0$ and should be modified
by subtracting off the zero mode contribution from the
right-hand-side.
\be
\vec{k}^2 \tilde{\phi}_{\vec{n}} = \frac{m}{M_{4}^2}  \left( e^{-i
  \vec{k}_n \vec{y} } - \delta_{\vec{n},0} \right)
\ee
This corresponds to \emph{not} integrating the zero mode out
of the effective theory and leaving $\tilde{\phi_0}$
unconstrained.  The corrected equation is
exactly the fourier transform of the modified Poisson equation,
Eq.\ (\ref{equ:potential_modified}), with the
$\vec{y}$-independent volume term corresponding to a Kronecker
delta which picks out the zero mode, thus justifying the volume subtraction.

We now
factorize out all dimensionful quantities as before to obtain
the following equation for the dimensionless potential
$\varphi$,
\be
\nabla^2 \varphi(\vec \theta) =
    \left( \delta^{(n)}(\vec \theta) - 1 \right)
\ee
which, away from all sources, is simply
\[
\left(\partial_1^2 + \cdots + \partial_n^2 \right)\varphi(\vec \theta) = - 1
\]
This sum of second derivatives, which determines the slow roll
parameter $\eta$,  will be minimized if we distribute it evenly
in all directions, giving $|\varphi''(\theta)| > 1 / n$.  This
constraint is a direct consequence of the modification in Eq.\ (\ref{equ:potential_modified}), and
results in a hard lower limit for $\eta$ that depends on the
number of branes and the number of extra dimensions, both of
which are $\mathcal{O}(1)$ quantities.

\section{Vanishing Gravity in Codimension 2}

The non-Newtonian nature of 2+1 D gravity provides a simple and
elegant way around these issues.  Point masses in infinite,
flat codimension 2 do not attract each other, but simply cut
out deficit angles that are proportional to their masses,
giving rise to conical spaces that are flat everywhere apart
from at the location of the masses \cite{Staruszkiewicz;1963}.
For completeness we review the argument for this below.

\subsection{Static point masses in 2+1 dimensions}
\label{sec:nonperturbativeflat}

We use the following ansatz for the line element (recall that 2
spatial dimensions are conformal to flat space):
\[
ds^2=dt^2-\omega(x,y)\left(dx^2+dy^2\right)
\]
For a single point mass at the origin of the space, the only
non-trivial component of Einstein's equation is:
\be\label{equ:einstein2+1}
-\f{2}\nabla^2\ln{\omega}=\frac{m}{M}\delta(\vec{r})
\ee
where $\nabla^2$ is the flat Laplacian, $m$ is the mass of the
source, and $M$ is the 2+1 D fundamental scale.   We solve this to
obtain
\[
\omega=\left|\vec{r}\right|^{-\frac{m}{\pi M}}
\]
Going to polar coordinates in the spatial dimensions gives a
metric:
\[
ds^2=dt^2-r^{-\frac{m}{\pi M}}\left(dr^2+r^2d\theta^2\right)
\]
but by making the coordinate redefinitions
\[
\rho=\f{1-\frac{m}{2\pi M}}r^{1-\frac{m}{2\pi M}}\;;\qquad
\theta'=\left(1-\frac{m}{2\pi M}\right) \theta
\]
we see that it indeed looks flat, with an additional
restriction on the polar angle:
\[
ds^2=dt^2-d\rho^2-\rho^2d\theta'^2\qquad\textrm{where}\qquad
0\le\theta'\le2\pi -\frac{m}{M}
\]
This is the metric for a circle with deficit angle $\delta =
m/M$ cut out, and the two boundaries of the deficit identified,
forming a cone with the point mass at its apex (see Fig.\
\ref{fig:cones}).
The static solution for multiple point masses is a space where
each one cuts out a deficit angle that is proportional to its
mass, with no gravitational force between them.

\FIGURE{
\psfrag{m}{$m$}
\psfrag{delta}{\!\!\!$\delta=\frac{m}{M}$}
\psfrag{m1}{$m_1$}
\psfrag{m2}{$m_2$}
\psfrag{delta1}{\!\!\!\!\!$\delta_1=\frac{m_1}{M}$}
\psfrag{delta2}{\!\!$\delta_2=\frac{m_2}{M}$}
\begin{tabular}{m{1.2in}m{0.1in}m{2.0in}m{1.9in}}
\includegraphics[height=1.2in]{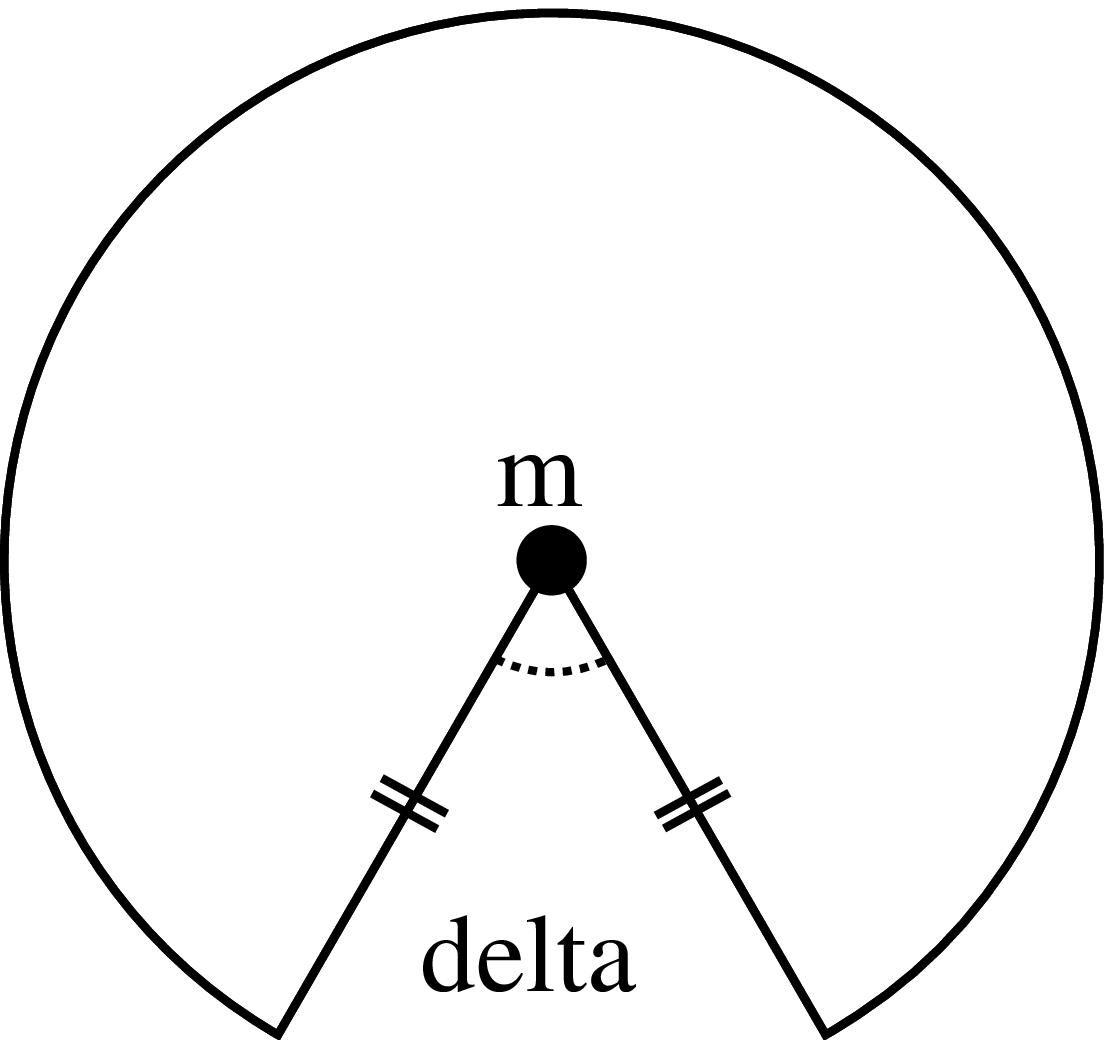}
&
\includegraphics[width=0.4in]{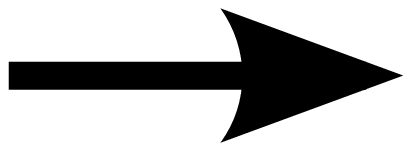}
&
\includegraphics[height=1.2in]{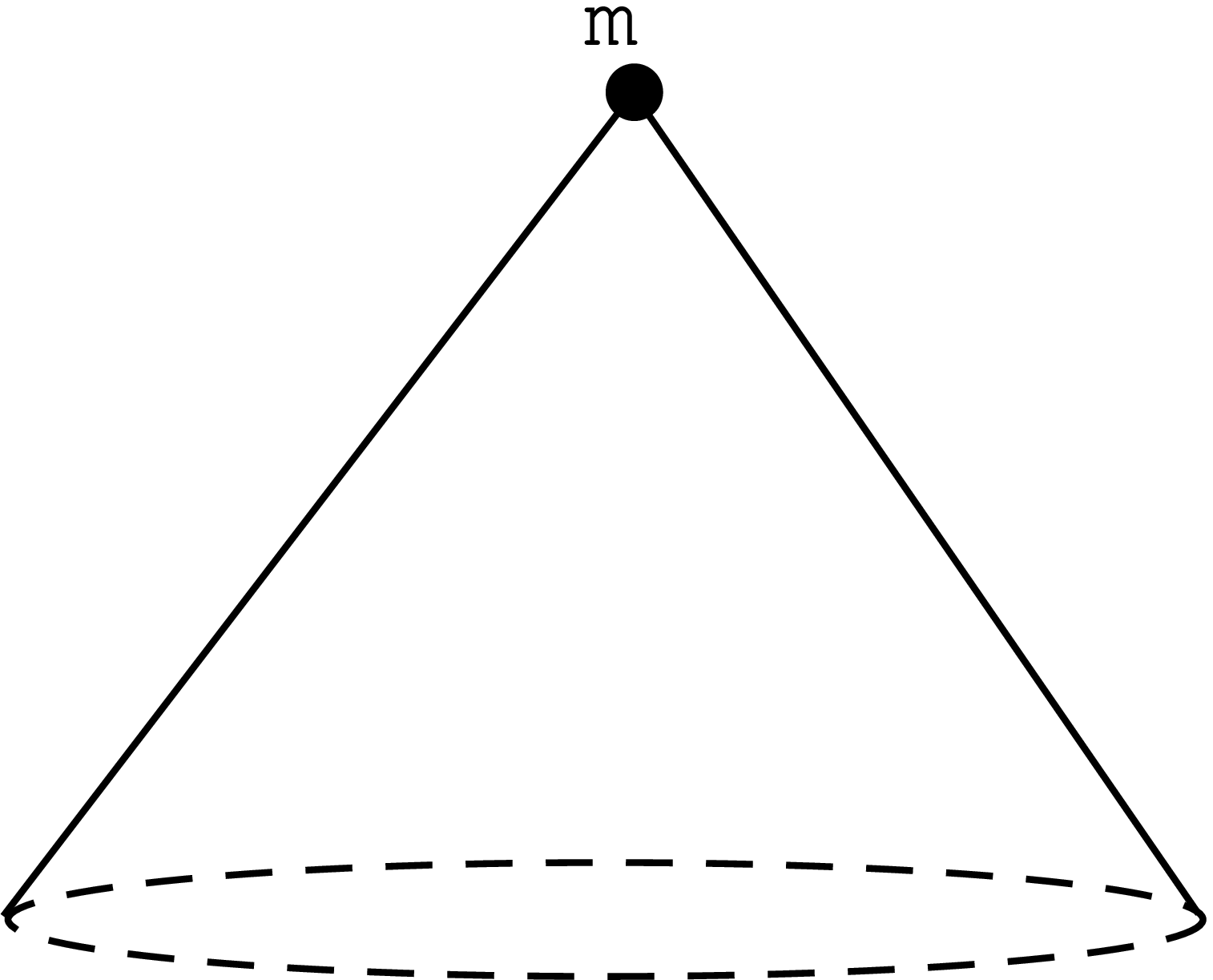}
&
\includegraphics[height=1.2in]{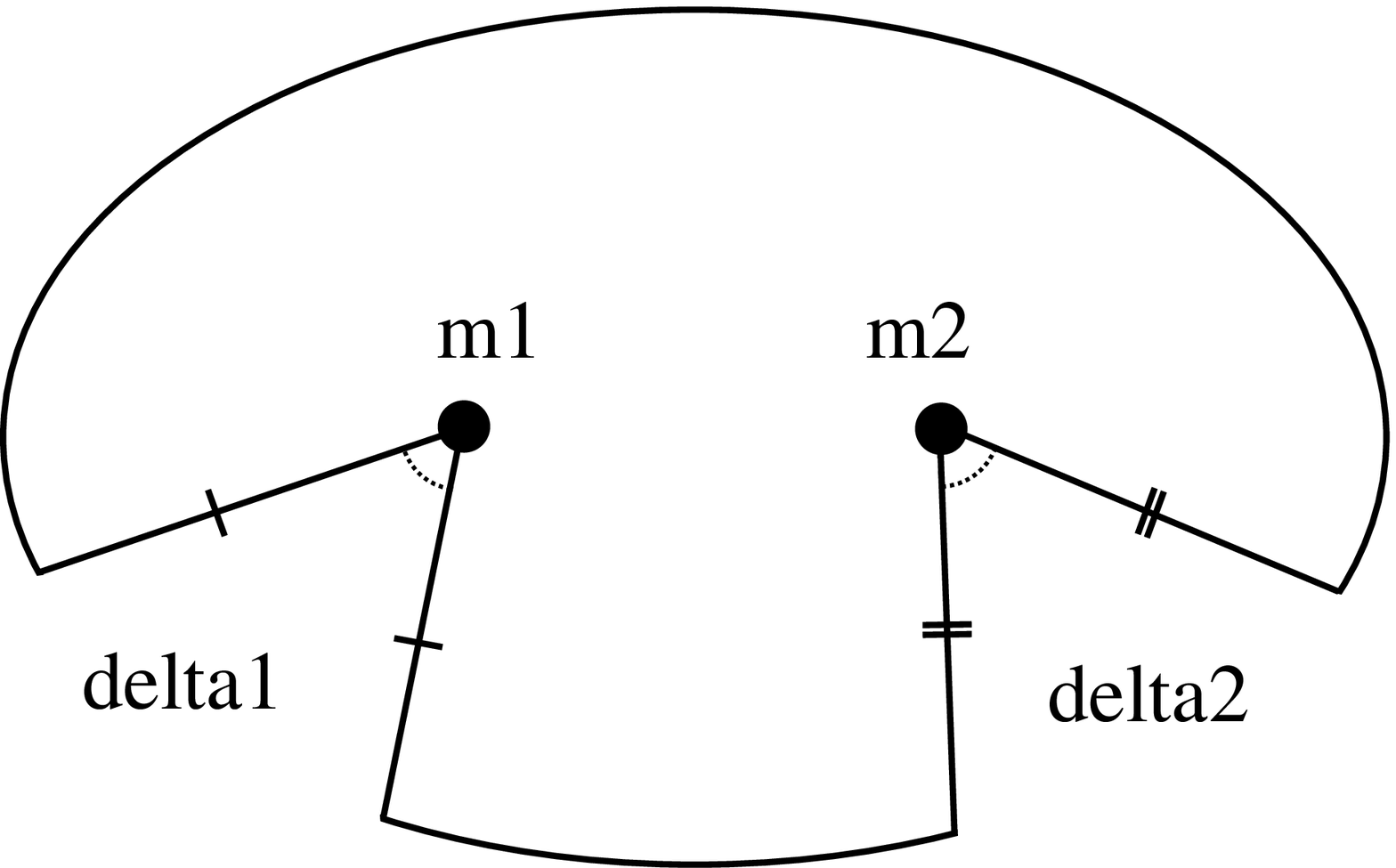}
\end{tabular}
\caption{A point mass in 2+1 dimensions cuts out a deficit
angle $\delta$ proportional to its mass (left), with the
two boundaries of the deficit identified, forming a cone
with the mass at its apex (center).  Multiple point masses
each cut out their own deficit angles, yielding as a static
solution a space that is flat everywhere except at the
positions of the sources (right).  Each mass has no effect
on any other - there is no gravitational force between
them.}
\label{fig:cones}
}

Since it similarly satisfies Eq.\
(\ref{equ:einstein2+1}) in the extra dimensions, a 3-brane in a
6-dimensional space also cuts out a deficit angle proportional to
its tension.  There is thus no gravitational
attraction between codimension-2 branes, circumventing altogether the
issue of a steep irreducible gravitational potential.

We explore this idea in detail below, studying the full 6D
problem in the perturbative limit, to leading order in the brane
sources. Our initial goal is to understand the
vanishing gravitational potential from the perspective of the
4D effective theory. We then use this knowledge to seek a realistic
inflationary model where the 2 extra dimensions are compactified and
stabilized.

\subsection{Perturbative codimension 2: flat space}
\label{sec:perturbativecodimension2}

We compute the effective potential by integrating out the 6D
graviton in the Einstein-Hilbert Lagrangian with a brane source
term (setting the 6D fundamental scale M=1 for simplicity)
\be\label{equ:6DflatLagrangian}
\mathcal{L}=-\sqrt{\left|g\right|}\left(\f{2}R + \f{2} T_{MN}H^{MN}\right)
\ee
where $T_{MN}$ is the energy-momentum tensor for a brane with
tension $f^4$ at position $\vec{y}=\vec{y}\,'$.  This is only
non-zero along the brane:
\be\label{equ:braneTmn}
T_{MN} = \left\{\begin{array}{cr}
                f^4 \, \delta\left(\vec{y}-\vec{y}\,'\right)\eta_{\mu \nu} &
                \ \ {\rm for \ \ M,N=0,\cdots,3} \\
                0 & {\rm otherwise}
                \end{array}\right. \ \ \ .
\ee
The potential between two brane sources is
\be
\label{eqn:source_propagator_source}
V=T^{(1)}_{MN} P^{MN \, OP} T^{(2)}_{OP}
\ee
with the graviton propagator $P^{MN \, OP}$ computed by
inverting the quadratic part of the gauge-fixed Lagrangian.
In de Donder gauge, the
propagator in $D$ flat space-time dimensions is
\be
 P^{MN \, OP} =
 \frac{1}{k^2} \left(
 \tfrac{1}{2} \, \eta^{MO} \eta^{NP} + \tfrac{1}{2} \, \eta^{MP} \eta^{NO}  - \tfrac{1}{D-2} \, \eta^{MN} \eta^{OP}
 \right)
\ee
For a $(d\!-\!1)$-brane this yields a potential:
\bea
V &\sim& \eta_{\mu \nu}
\left(
 \tfrac{1}{2} \, \eta^{JM} \eta^{KN} + \tfrac{1}{2} \, \eta^{JN} \eta^{KM}  - \tfrac{1}{D-2} \, \eta^{JK} \eta^{MN}
 \right)
\eta_{\sigma \rho}\nonumber \\
&=& \frac{1}{2} d + \frac{1}{2} d -  \frac{d^2}{D-2} \\
&=& 0 \qquad \textrm{for } D=d+2 \qquad \textrm{(codimension 2)} \nonumber
\eea
By compactifying this space on a flat 2-torus one
can understand the mechanism by which this cancellation occurs in the 4D effective
theory.  We sketch below the argument showing that it stems from a mode-by-mode cancellation
between Kaluza-Klein (KK) modes of the 4D graviton tower and
those of the 4D radion tower.

The 6D metric perturbation $H_{MN}$ around a flat background can be
parametrized as follows:
 \be
H_{MN}  = \left( {\begin{array}{*{20}c}
   {h_{\mu \nu} - 2 \Phi \eta _{\mu \nu }} & {V_{\mu n} }  \\
   {V_{\nu m} } & {\phi _{(mn)} + 2 \Phi \eta_{mn}}  \\
 \end{array} } \right)
\ee
where parentheses indicate trace-subtracted indices, e.g.
$\eta_{mn}\phi^{(mn)}=0$.

Not all these are physical degrees of freedom
since there are gauge redundancies: specifically 6 general
coordinate transformation degrees of freedom which need to be fixed.
In this simple scenario it is easy to choose a gauge that decouples
all fields at quadratic order:
\be
  \partial ^m V_{\mu m}  = 0 \qquad \qquad \partial
  \phi_{(mn)}  +
  4\partial _n \Phi = 0
\ee

Canonically normalizing the independent physical fields, we encounter a
tower of graviton and radion KK modes of equal mass $m_{(i,j)}$,
with propagators $P_{(i,j)}^{\mu\nu\,\rho\sigma}$ and $P_{(i,j)}$
respectively\footnote{There is also a tower of vector modes that does
  not couple to the branes or mix with other fields.}.
We can now compute the contribution to the potential between branes
in the limit of conserved sources, due to integrating out the graviton
and radion towers.  For each mode $(i,j)$ we find a potential
\begin{eqnarray}
  V_{\left(i,j\right)} &\sim& \left( \eta _{\mu \nu }P_{(i,j)}^{\mu \nu \sigma \rho } \eta _{\sigma \rho }+ \frac{2}
{\sqrt 3 }P_{(i,j)}  \frac{2}
{\sqrt 3 }\right)\\
   &\propto&  - \frac{4}
{3} + \frac{4}
{3} =0 \nonumber
\end{eqnarray}
i.e.\ at \emph{each} level the radion KK mode provides an
attractive force that is exactly equal to the repulsive force
due to the graviton KK mode: mode by mode the force cancels.

With no gravitational attraction between them, 3-branes in
infinite 6D are stable.  However, they source a linear
potential for the (zero mode) radion in the effective theory,
which sets the size of the compact space, causing it to run
away to infinity.  A reliable, calculable theory of brane
inflation must provide a mechanism to stabilize all zero modes
that couple to the inflaton field~\cite{Kachru:2003sx}.

The first example of a successful stabilization method for
compact extra dimensions was due to Freund and
Rubin~\cite{Freund:1980xh}, who employed a combination of a bulk
cosmological constant and an extra-dimensional magnetic flux to
give a mass to the radion mode, fixing the radius of the $n$-sphere.
We use this mechanism to stabilize 2 extra dimensions compactified on a sphere
\cite{RandjbarDaemi:1982hi}, and compute the 4D effective potential between
static probe branes in this $\rm{Minkowski}_4\times S_2$ background
(see Fig.\ \ref{fig:compact2sphere}).

\FIGURE{
\psfrag{Fmn}[]{\;\;$F_{mn}$}
\psfrag{L6}{$\Lambda$}
\psfrag{B}[]{$B$}
\includegraphics[height=1.3in]{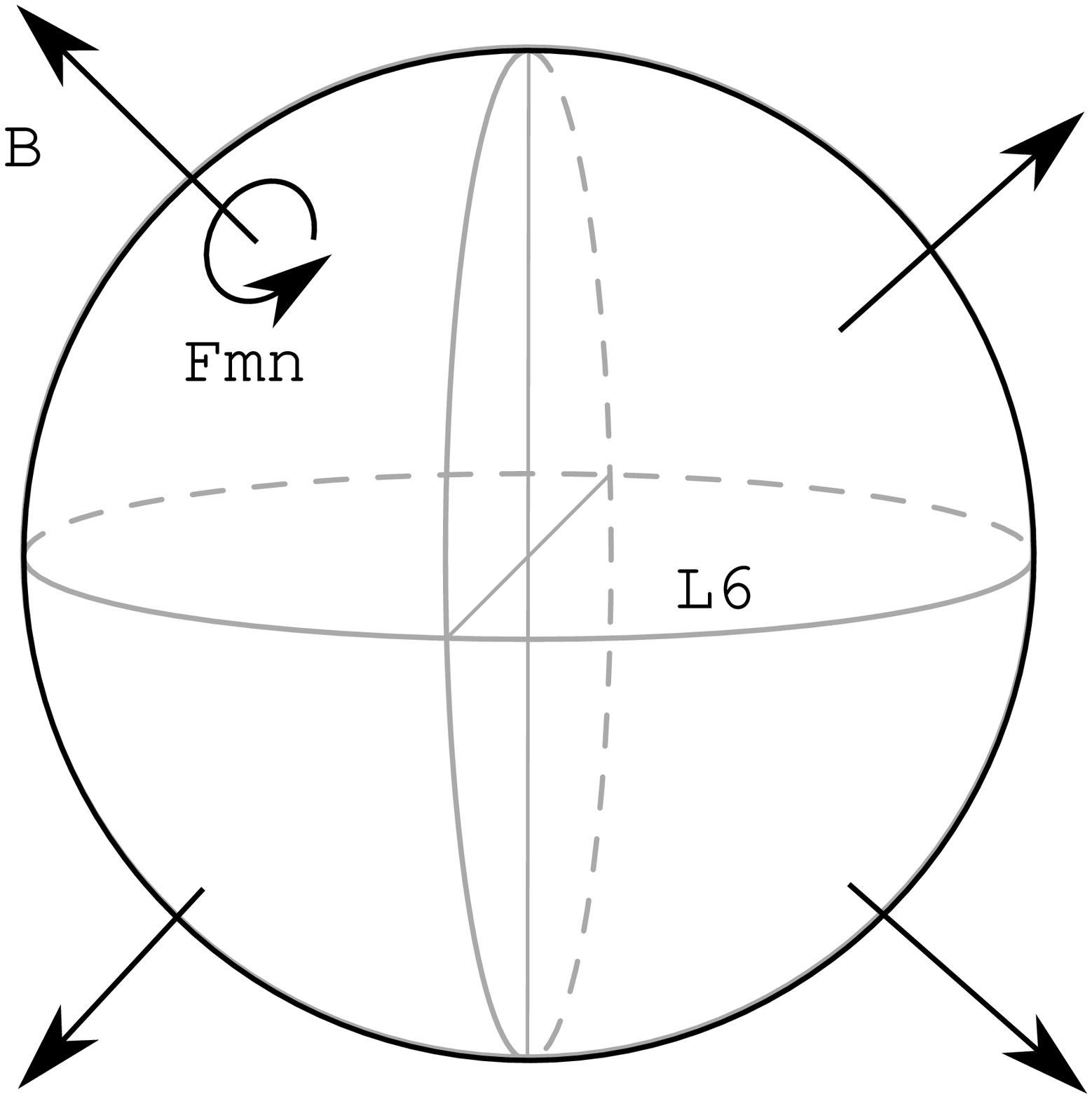}
\qquad\qquad\qquad
\includegraphics[height=1.3in]{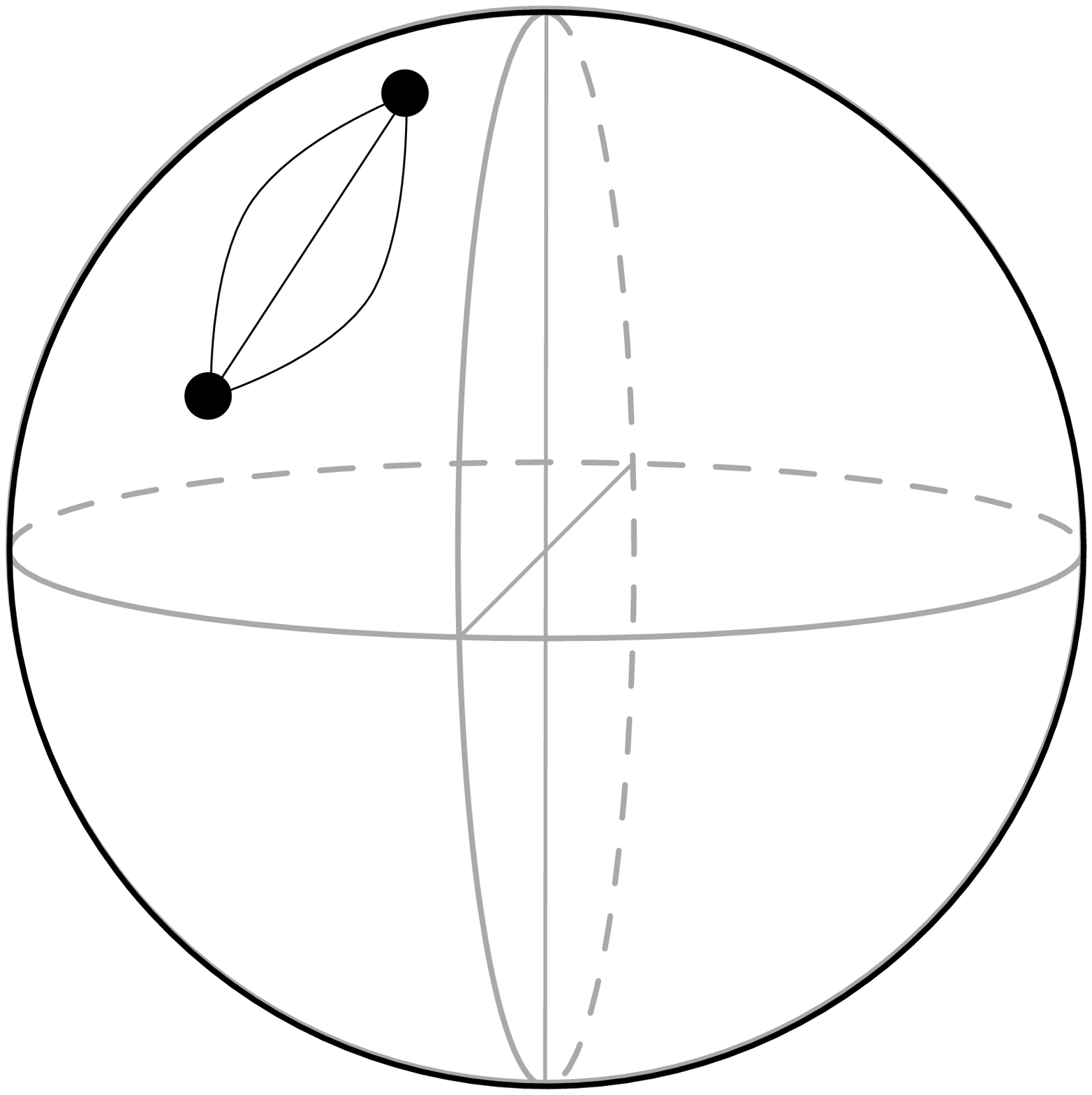}
\caption{Compact 2-sphere stabilized by combination of 6D
  cosmological constant and extra-dimensional magnetic
  flux (left); two probe-branes on stable 2-sphere
  (right).}
\label{fig:compact2sphere}
}

\subsection{Perturbative codimension 2: compact 2-sphere}
\label{sec:FR}

The compactified background metric takes the form of a
2-sphere orthogonal to infinite 4D Minkowski space:
\[
ds^2 = \eta_{\mu \nu} dx^\mu dx^\nu - r^2 \left( d\theta + \sin^2\theta \ d\phi \right)
\]
In addition to the Einstein-Hilbert term, the stabilized Lagrangian contains a 6D cosmological constant and an
electromagnetic field strength tensor:
\begin{equation}
\label{eqn:gravity_em_lagrangian}
\mathcal{L}=-\sqrt{\left|g\right|}\left(\f{2} R+\Lambda
+\f{4} F_{MN}F^{MN}
\right)
\end{equation}
Expanding this to linear order in perturbations $H_{MN}$ about
the background metric $\gb_{MN}$, gives Einstein's equation for
the background:
\begin{equation}\label{equ:einstein}
\left( \Rb_{MN}-\f{2}\gb_{MN}\Rb \right)=\gb_{MN}\Lambda+\f{4}\gb_{MN}\Fb_{OP}\Fb^{OP}-
\Fb_{MO}\Fb_{N}^{\ O}
\end{equation}%
In general the background electromagnetic field strength must
also be perturbed in order to obtain a consistent solution for
Einstein's equation with arbitrary
sources.  We confirm in Appendix \ref{app:flux} that
the total flux through the sphere is unaffected by the presence of
such a perturbation, which is defined as follows:
\[
F_{MN}=\Fb_{MN}+f_{MN}
\]
and can be written
in terms of a dimensionless gauge potential $b_N$, and a constant, $B$:
\[
B \, f_{MN}=\nabla_M b_N-\nabla_N b_M = \nb_M b_N-\nb_N b_M
\]
Setting the linear coefficient of $b_N$ in the Lagrangian to
zero gives Maxwell's equation $\nb_M \Fb^{MN}=0$, one possible
solution to which is
non-zero only in the extra-dimensional magnetic components:
\be
\bar F_{MN}=\left\{\begin{array}{cr}
                     0 & \ \ {\rm for \ M,N=0,\cdots,3}\\
                     B \, \epsilon_{mn} & \ \ {\rm otherwise}
                   \end{array} \right.
\ee
where $\epsilon_{mn}$, the extra-dimensional spherical
surface-area element, is related to the totally antisymmetric
tensor $\varepsilon_{mn}$ by $\epsilon_{mn} = \sqrt{g_{mn}}
\varepsilon_{mn}$ (with $\varepsilon_{45}=1$).  The constant
$B$ is now seen to be the strength of the magnetic field.

The 6D and 4D trace of Einstein's equations relate the
magnetic field $B$ and cosmological constant $\Lambda$ to the
radius of the sphere $r$:
\be\label{equ:tuning}
\Lambda =  \frac{1}{4}\Fb_{mn}\Fb^{mn} = \frac{1}{2} B^2 =
-\frac{1}{4}\Rb  = \frac{1}{2r^2}
\ee
where $\Rb$ is the background Ricci scalar.  This fixes all
components of the Riemann Tensor in this background,
\be\label{equ:bkgsimplify}
\Rb_{MNOP} = -\Fb_{MN}\Fb_{OP}  \\
\ee
and sets to zero the sum of the constant background terms in the Lagrangian
(\ref{eqn:gravity_em_lagrangian}), as expected for an
effective 4D Minkowski space\footnote{If the cosmological
constant and magnetic field were not perfectly balanced, we
would have an inflating de Sitter space instead of flat
Minkowski. The imbalance between the cosmological constant and
the magnetic field would determine the Hubble parameter of the
4D Minkowski space~\cite{Martin:2004wp}.}.

Including a source term for the branes, which have
energy-momentum tensor $T_{MN}$ given by Eq.\
(\ref{equ:braneTmn}), and simplifying using the background
solution (Eqs. (\ref{equ:einstein}) and
(\ref{equ:bkgsimplify})), the Lagrangian to quadratic order in
perturbations is:
\begin{eqnarray}\label{equ:quad}
\mathcal{L}&=&\f{4}H^{MN}\left[\f{2}\left(g_{MN}g_{OP}-g_{MO}g_{NP}\right)\nabla^2+g_{NP}\left(\nabla_M\nabla_O+R_{MO}\right)-g_{MN}\nabla_O\nabla_P\right]H^{OP}\nonumber\\
&+&\f{4}H^{MN}\left[4 F_{MO}f_N^{\ O}-g_{MN}F_{OP}f^{OP}\right]-\f{4}f_{MN}f^{MN} - \f{2} H^{MN} T_{MN}
\end{eqnarray}
where we have omitted the bars
denoting background quantities for simplicity.
This Lagrangian yields the following linear equations of motion for $H^{MN}$ and
$b^N$:
\begin{eqnarray}\label{equ:eomh}
g_{MN}\big(\nabla^2 \!H - \nabla_O\!\nabla_P H^{OP}\!\!-F_{MN}f^{MN}\big)
  -\nabla_M\!\nabla_N H -\nabla^2 \!H_{MN}\hspace{1.2in}&&\\
  +\Big[\left(\nabla_M\!\nabla_O + R_{MO}\right) \!H_{N}^{\ O}+ 2 F_{MO}f_{N}^{\ O}\!+ \left(m\!\!\leftrightarrow\!
  n\right)\Big]&=&2 \; T_{MN}\nonumber\\
-\frac{1}{2}F^{MN}\nabla_M H + \nabla_M H_O^{\ M}F^{ON}-\nabla_M
H_O^{\ N}F^{OM}- \nabla_M\nabla^M b^N+\nabla_M\nabla^N b^M&=&0
\ \ \nonumber .
\end{eqnarray}
We KK reduce by parameterizing the metric perturbations as follows:
\be\label{equ:parah}
H_{MN}  = \left( {\begin{array}{*{20}c}
   {h_{(\mu \nu)}  + 2 \Psi \eta _{\mu \nu } } & {V_{\mu n} }  \\
   {V_{\nu m} } & {\phi _{(mn)} + 2 \Phi g_{mn} }  \\
 \end{array} } \right)
\ee
Note that we separate the 4D component into traceless part and trace
part and do not make a Weyl transformation, for reasons that will
become clear later. The gauge field perturbation
$b_M$ is decomposed as\footnote{In general there is also a
  divergenceless harmonic form
$\beta$, but this is automatically zero on a sphere (``You
cannot comb the hair on a sphere'').}
\bea\label{equ:parab} b_M  = \left(
{\begin{array}{*{10}c}
   b_\mu  \\
   \epsilon_{km} \nabla^k b + \nabla_m b'  \\
 \end{array} } \right)
\eea

Due to the non-zero curvature and magnetic flux, there is no viable
gauge in which all the physical fields in this particular
parametrization are decoupled.
We gauge fix as follows, making sure to also fix the
additional $U(1)$ gauge freedom:
\be\label{equ:gauge}
\nabla_n V^{n\mu} = 0
\qquad \qquad
\nabla_n \phi^{(mn)}=0
\qquad \qquad
\nabla_n b^n = 0
\ee
which sets to zero the following:
\be
\phi^{(mn)} = 0
\qquad \qquad
b' = 0.
\ee

We can parametrize the extra-dimensional dependence of the 4D
scalar fields using scalar spherical harmonics
$Y_{\ell,m}(\theta,\phi)$.  Then we can perform a separation of
variables as follows:
\be
\nabla_M\nabla^M\psi\left(x^\mu,\theta,\phi\right)=\left(\Box+\triangle\right)
\psi\left(x^\mu\right)Y\left(\theta,\phi\right)
\ee
for a general field $\psi$, where $\Box=\partial_\mu \partial^\mu$, and $\triangle
Y=\nabla_m\nabla^m Y = r^{-2}\ell\left(\ell+1\right)$.

We compute the potential by integrating out the massive scalar
KK modes for static brane sources ($p_\mu p^\mu \ll m^2$) by inverting the scalar
mass matrix. The quadratic mass mixing at each KK level in the
Lagrangian (suppressing all spherical harmonic
$\left(\ell,m\right)$ indices for simplicity) takes the form
\be
\label{equ:massmix}
\mathcal{L} \supset r^{-2} \ell(\ell+1)
\left( \Psi \quad \Phi \quad b \right)
\cdot
\left(\begin{array}{ccc}
6 & 2 & -2 \\
2 & -\frac{2}{\ell(\ell+1)} & 1\\
-2 & 1 & -\frac{\ell(\ell+1)}{2}\\
\end{array}\right)
\cdot
\left(\begin{array}{c}
\Psi  \\
\Phi \\
b \\
\end{array}\right)
\ee
For $\ell \ge 2$ this mass matrix is invertible, giving rise to a
potential
\be
V  \sim \left(\begin{array}{ccc}1&0&0\end{array}\right)\cdot
\left(\begin{array}{ccc}
6 & 2 & -2 \\
2 & -\frac{2}{\ell(\ell+1)} & 1\\
-2 & 1 & -\frac{\ell(\ell+1)}{2}\\
\end{array}\right)^{-1}
\cdot\left(\begin{array}{c}1\\0\\0\end{array}\right)=0\ \ .
\ee

For $\ell\!=\!1$ however, the mass matrix has a zero
eigenvalue. This is not due to the presence of a physical massless
field, but rather due to an additional gauge redundancy, corresponding
to conformal diffeomorphisms on the sphere \cite{Martin:2004wp,
vanNieuwenhuizen:1984iz}, which needs to be fixed.  The static
quadratic Lagrangian for $\ell=1$ is only a function of $\Psi$
and the combination $\chi=\Phi-b$, and is completely
independent of the orthogonal combination $\chi_\perp=\Phi+b$,
which is a pure gauge mode and drops out.  The mass matrix in the
$\{\Psi,\chi\}$ basis can be inverted to yield a potential between
two branes of equal tension $f^4$, separated by angle $\theta$
on the sphere, of
\be\label{equ:leq1pot}
V = \frac{2 f^8}{5 M^4} \sum_{m=-1}^{+1}  \overline { Y_{1m}}(0,0) Y_{1m}(\theta, \phi)
  =  \frac{3 f^8}{10 \pi M^4} \cos(\theta)
\ee

This \emph{repulsive} potential between the probe branes gives
a dimensionless shape function $\varphi\left(\theta\right)$
that goes like $\cos\left(\theta\right)$.  Hence $\eta\sim
-\cos\left(\theta\right)$
which is $O\left(1\right)$ for arbitrary positions of the branes
on the sphere.  Furthermore, although $\eta(\pi/2)=0$,
the force between the branes is maximal and repulsive, at this point,
yielding an unstable configuration.
These results are confirmed by analyzing the linearized equations
of motion for the KK modes in Appendix \ref{app:KKeom}.

We want to emphasize that, from the perspective of the 4D
effective theory, the simplicity of this result is rather
surprising.  We saw above that the absence of a gravitational
potential between two codimension-2 sources in a flat
background could be perturbatively attributed to a mode-by-mode
cancelation between the graviton KK tower and the radion KK.  One might expect
that stabilizing the space by giving a mass to the radion mode
while keeping the graviton massless would shift the masses of
the entire radion KK tower while leaving the graviton tower
untouched, spoiling the cancelation at every KK
level.  This also what one would naively expect from
examining the quadratic Lagrangian, Eq.\ (\ref{equ:quad}): the
background curvature $R_{MO}$, having only extra-dimensional
components, acts like an extra contribution to the  mass of the
radion tower, and has no effect on the graviton tower.
However, mixing with the flux perturbation can be thought of as
undoing this effect
for $\ell\ge 2$, leaving only the contribution from the
$\ell=1$ mode for arbitrary brane positions.  The exact
mechanism by which this cancellation takes place can be
elucidated only by diagonalizing the full quadratic Lagrangian,
including kinetic terms, in order to identify the real
propagating degrees of freedom.  Details can be found in a
follow-up paper \cite{Gallicchio:branesymmetrybreaking}.

The striking simplicity of the resulting potential allows for a
similarly simple solution: a spherical background that has
antipodal points identified is also a solution to the
background equations, but one with all odd KK modes projected
out, excluding the problematic $\ell=1$ mode from the spectrum.   This space can be thought of as the
real projective space $P_2$ (see Fig.\ \ref{fig:RP2}, left
panel), or equivalently as the entire $S_2$ with brane sources
appearing in antipodal pairs (Fig.\ \ref{fig:RP2},
right panel).\footnote{This is distinct from the
  more common orbifold $\mathbb{Z}_2$ projection across the
equator of the sphere.}

\FIGURE{
  \begin{tabular}{cc}
\parbox[t]{2in}{\includegraphics[width=1.5in]{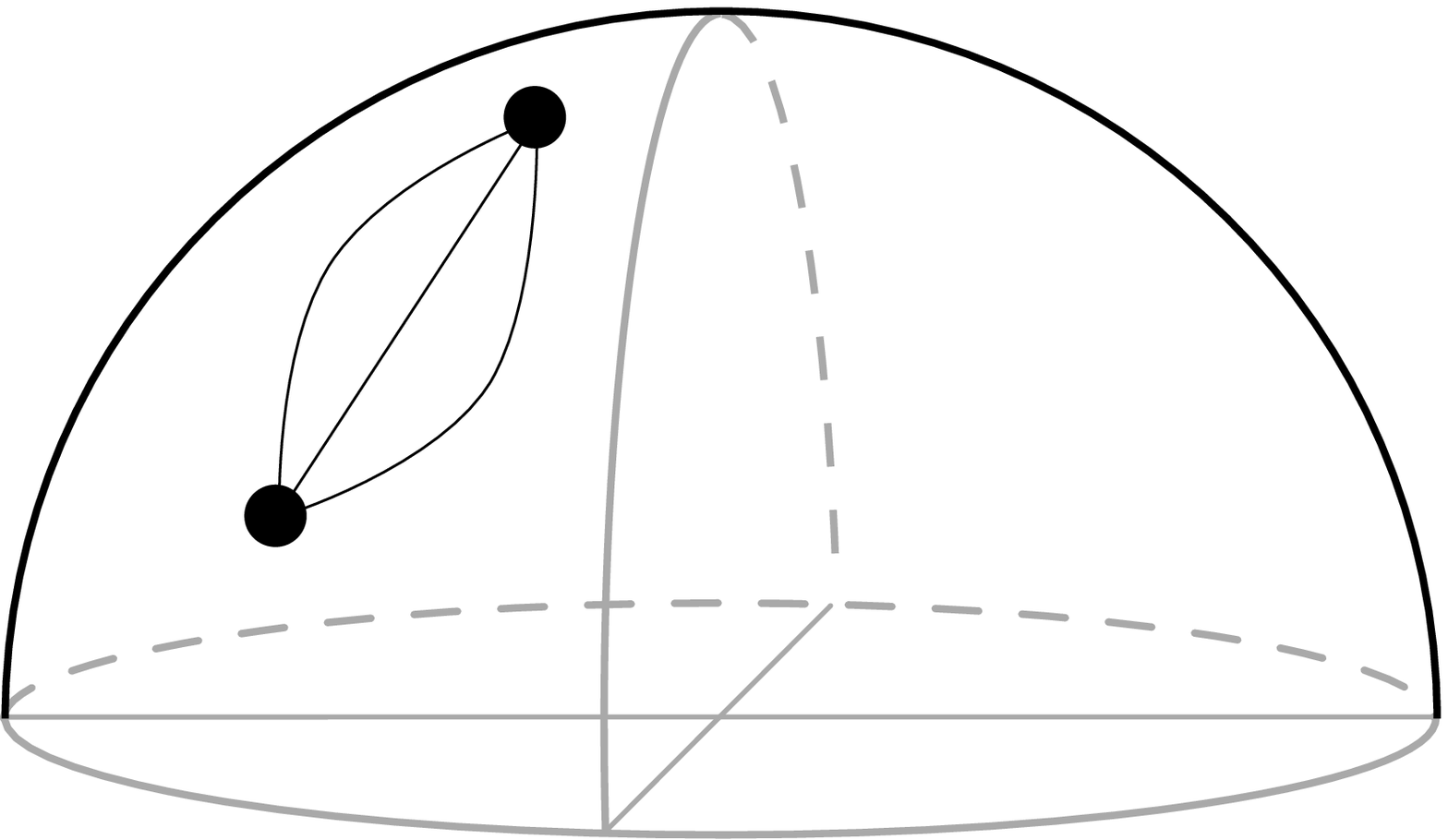}}
&
\parbox{2in}{\includegraphics[width=1.5in]{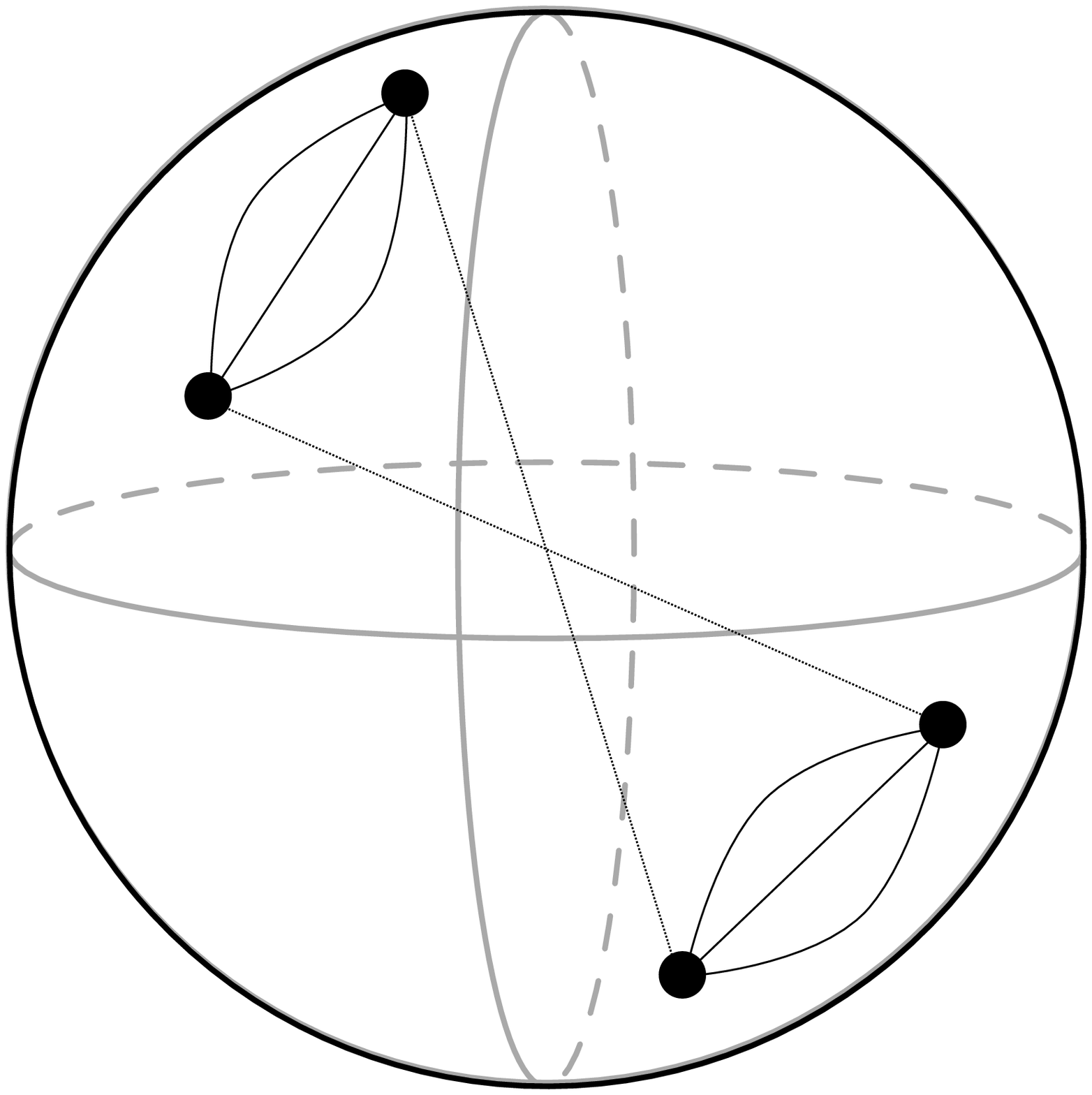}}
\end{tabular}
\caption{A single brane on the projective space $P_2$ (left) can be
equivalently thought of as an antipodal pair of branes on $S_2$
(right).
  This setup projects out all odd KK modes, setting to zero the
  gravitational potential for arbitrary brane positions.}
\label{fig:RP2}
}

It is evident from considering the lower-dimensional effective theory that
the brane tension here plays the role of the 4D vacuum energy: taking the
tension to zero results in a static space that does not inflate.
This relationship between the brane
tension and the 4D cosmological constant is quantified, in the
non-perturbative limit with finite-tension branes, in Appendix
\ref{app:non-perturbative}. It is natural to ask whether
our result of zero gravitational potential still applies in this limit.  The
following intuitive argument\footnote{due to Markus Luty} suggests
that it does.  Dialing up the
brane tension, in the antipodal-pair picture
each pair of branes
cuts out a slice of the sphere of angular size proportional to
its tension. These slices can be chosen to have no overlap with
each other, in which case it is clear that the first pair of
branes has no effect on the second pair, and the gravitational
potential is zero (see Fig.\ \ref{fig:two_wedges}). In addition, the
fact that the potential before the projection was repulsive ensures
the stability of any finite-tension setup with maximally separated branes,
e.g.\ the football configuration in \cite{Carroll:2003db}.

\FIGURE{
\qquad \qquad \qquad \qquad
\includegraphics[width=1.5in]{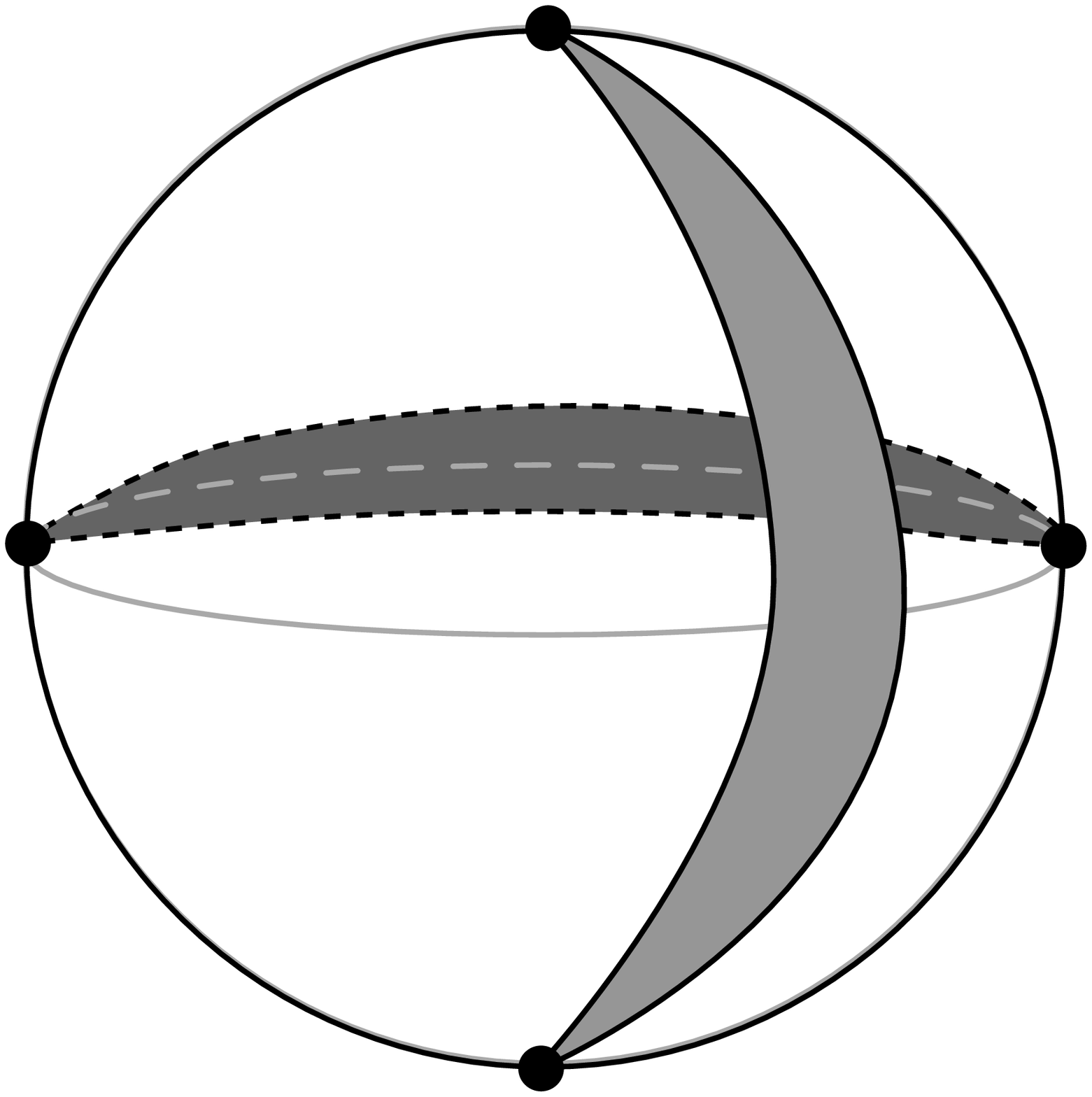}
\qquad \qquad \qquad \qquad
\caption{Two antipodal pairs of branes with finite tension each cut out
their own deficit angle.  These regions can be chosen not to overlap, implying
zero (gravitational) interaction.}
\label{fig:two_wedges}
}

One might still be concerned about the stability of our setup
being jeopardized by the presence of three further massless
modes, corresponding to orthogonal fluctuations of the branes
in the extra dimensions.  However these are `eaten' by an
$SU(2)$s-worth of gauge bosons (originating in the off-diagonal
elements of the 6D graviton)
\cite{Parameswaran:2006db,Dobado:2000gr}, which then get a mass
$m_{SU(2)}\sim\sqrt{\delta}/r$.

\begin{wrapfigure}{l}{0.3\textwidth}
\psfrag{E}{$E$}
\psfrag{M4}{$M_4$}
\psfrag{M}{$M=\sqrt{\frac{M_4}{r}}$}
\psfrag{Lambda6}{$\sqrt[6]{\Lambda_{6D}}\sim\sqrt[3]{B}\sim\sqrt[3]{\frac{M_4}{r}}$}
\psfrag{f}{$f=\sqrt[4]{\delta}M\sim\sqrt[4]{\Lambda_{4D}}$}
\psfrag{size}{$\frac{1}{r}$}
\psfrag{H}{$H\sim m_{SU(2)}\sim\frac{\sqrt{\delta}}{r}$}
\includegraphics[width=0.7in]{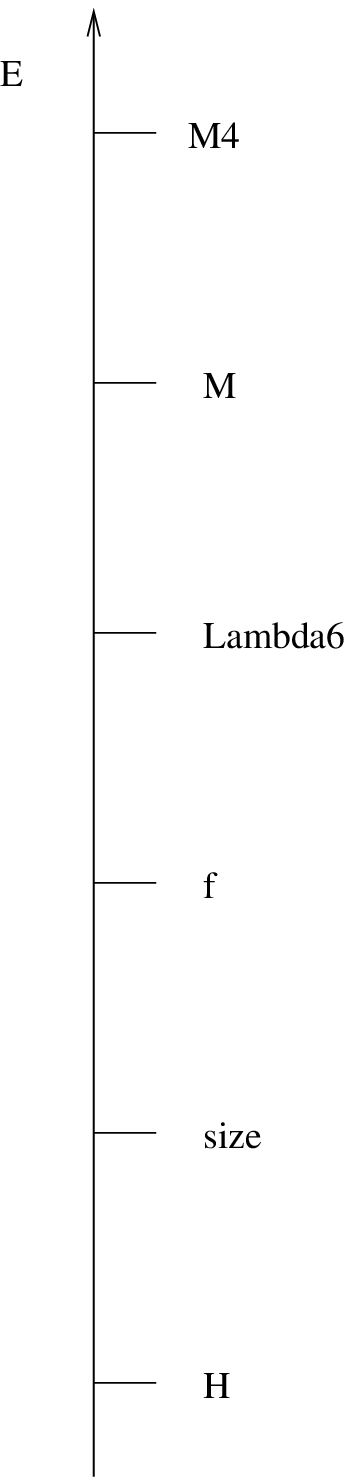}
\label{fig:scales}
\end{wrapfigure}

The different scales in this construction and their relationship to
each other can be seen on the left.  Note that the
brane tension $f^4$, is not set, and can vary anywhere in the range
$1/r\le f\ll M$.

Having eliminated the troublesome irreducible gravitational
interaction between two brane sources, we
envision various scenarios to generate a slow-roll
potential. We could couple the branes to a massive bulk scalar, for
example, with a
judicious choice of parameters to ensure flatness.  A less ad-hoc
method would be to use the Casimir potential between the branes
(quantum corrections to the potential due to loops of massive KK
modes). On purely dimensional grounds we
expect this to scale like $f^4/\left(r M\right)^4$
~\cite{Sundrum:1998ns}, which is relatively flat.  A full
computation of the one-loop effective potential
is beyond the scope of this paper, but will be addressed in
a future publication \cite{Gallicchio:branesymmetrybreaking}.

\section{Conclusion}

In this paper we tackle a problem that plagues simple
extra-dimensional models of brane inflation: the presence of an
irreducible gravitational component to the potential between
branes making it generically too steep to satisfy inflationary
slow-roll conditions. Attempts to solve this problem have usually invoked the
framework of string theory, at the cost of introducing various
light moduli that couple to the inflaton
and need to be stabilized.  Our strategy is to capitalize on
the special properties of gravity in codimension 2, where there
is no gravitational potential between point sources in flat space.
We focus on a setup with probe 3-branes perturbing a 2-sphere
background, with the radius of the sphere stabilized by a
magnetic flux and a 6D cosmological constant.
We find that the only contribution to the inter-brane
potential in the effective theory is due to the
exchange of level-1 KK modes, and is easily eliminated by
projecting out half the sphere, removing all odd modes from the theory.
Then a realistic model for codimension-2 brane inflation requires some
other source of a small inflationary potential.  This could be added
by hand, by coupling the branes to a massive bulk scalar, for example,
or it could already be present, in the form of small one-loop casimir corrections
to the gravitational potential due to massive KK modes.

We do not discuss brane collisions, graceful
exit or reheating. We have no reason to think that
this model will yield results that are any different from those already considered in the
literature, including warped inflation models (see
e.g. \cite{Kofman:2005yz}).  We leave an in-depth analysis for future work.

Although initially surprising, with hindsight we do not expect the
 near-cancellation of the
potential to be specific to the case of a spherical background.
Extrapolating from the perturbative 6D flat-space result, we would
predict the same outcome, for all higher modes, in
any background that asymptotes to flat space,
 although stabilizing and computing the potential might
prove more difficult in other configurations.  Hence we
believe that codimension-2 setups generically alleviate the problem of
a too-steep gravitational potential, perhaps bringing us one step
closer to finding a workable framework within which to study brane inflation.

\section*{Acknowledgements}
We are indebted to Nima Arkani-Hamed who suggested this
project, and guided us through the various difficulties that
arose along the way. Many thanks also to Gia Dvali, Raphael Flauger, Markus Luty,
John
Mason, Raman Sundrum and Toby Wiseman for helpful discussions,
and to KITP, where this project was revived, for its
hospitality. Fermilab is operated by Fermi Research Alliance,
LLC under Contract No. DE-AC02-07CH11359 with the United States
Department of Energy.  This research was supported in part by
the National Science Foundation under Grant No. PHY05-51164 and
the Bosack \& Kruger Foundation.

\appendix

\section{Perturbing the Flux}\label{app:flux}
The total flux through the sphere is quantized, and
should be constant at all orders even after taking into account
perturbations.

\begin{equation}
{\rm Flux}=\frac{1}{2}\int{\varepsilon^{mn}F_{mn} \;d\theta\,d\phi}
\end{equation}
for $\varepsilon_{45}=1$.  Perturb around background:
$F_{mn}=B\epsilon_{mn}+f_{mn}$, where $\epsilon_{mn}=\sqrt{g_{mn}}\,\varepsilon_{mn}$
\begin{eqnarray}
{\rm Flux} &=& \int{\left[B + \frac{1}{2}\epsilon^{mn}\left(\nabla_mb_n-\nabla_nb_m\right)
\right]\sqrt{g_{mn}}\,d\theta\,d\phi}\nonumber\\
&=& \int{\left(B + \triangle b\right)d^2S}\\
&=& B\times \left({\rm Surface \ area}\right) + \int{\triangle b \; d^2S}\nonumber
\end{eqnarray}
Decomposing $b$ into scalar spherical harmonics we see that
perturbations trivially give zero contribution to the flux to
all orders.

\section{KK Equations of Motion}\label{app:KKeom}

As a check of the non-intuitive results in Section \ref{sec:FR}, we
compute the linearised equations of motion for the graviton KK modes.
We roughly follow the procedure outlined in
\cite{Martin:2004wp}, although one must be wary when comparing
results since the latter uses the `mostly plus' sign
convention and a Weyl rescaled graviton trace $\Psi$.

Expressing the different components of the linearized equation
of motion for the graviton Eq.\ (\ref{equ:eomh}) in terms of the fields in Eqs.
(\ref{equ:parah}) and (\ref{equ:parab}), and using the gauge
conditions specified in Eq.\ (\ref{equ:gauge}), we obtain
\bea
\label{equ:solh}
\mu\nu:\ \ \ \eta_{\mu\nu}\left(4\Box\Psi+4\Box\Phi+6\triangle\Psi+2\triangle\Phi-
\partial_\alpha\partial_\beta h^{(\alpha\beta)}-2\triangle
b\right)-4\partial_\mu\partial_\nu\left(\Psi+\Phi\right)\hspace{0.2in}&&\nonumber\\
-\left(\Box+
\triangle\right)h_{(\mu\nu)}+\partial_\mu\partial^\sigma
h_{(\sigma\nu)}+ \partial_\nu\partial^\sigma h_{(\sigma\mu)}&=& 2
T_{\mu\nu}\nonumber\\
\mu n: \ \ -\left(\Box+\triangle\right)V_{\mu n}+\partial_\mu\partial^\nu V_{\nu
  n}-2B^2 V_{\mu n}+2\epsilon_{mn}\nabla^m b_\mu\hspace{1.7in}&&\\
+\nabla_n\left[\partial_\mu\left(-6\Psi-2\Phi+2 b\right)
  +\partial^\nu h_{(\mu\nu)}\right]&=&0\nonumber\\
mn: \ \ g_{mn}\left(6\Box\Psi+8\triangle\Psi+2\Box\Phi-4B^2\Phi-\partial_\mu\partial_\nu
h^{(\mu\nu)}+2\triangle b\right)-8\nabla_m\nabla_n\Psi\hspace{0.5in}&&\nonumber\\
+\nabla_m\partial^\mu V_{\mu n}+\nabla_n\partial^\mu V_{\mu m}&=&0\nonumber
\eea
Next, we isolate the extra-dimensional dependence as scalar,
vector and trace-subtracted tensor spherical harmonics
$\left\{Y,\;\nabla_m Y,\; Y_{(mn)}=\left(\nabla_m\nabla_n
-\frac{1}{2}g_{mn}\triangle\right)Y\right\}$, which are
linearly independent of each other.  After some manipulation,
the equations for scalar combinations
$\left\{\partial_\mu\partial_\nu
h^{(\mu\nu)},\Psi,\Phi,b\right\}$ decouple from the others to
give (suppressing $(l,m)$ spherical harmonic indices for
simplicity):
\bea
\left(3\Box\Psi+3\Box\Phi+6\triangle\Psi+2\triangle\Phi-\f{2}\partial_\mu\partial_\nu h^{(\mu\nu)}-2\triangle b\right)Y &=&
\left( 2 T^{\mu}_{\mu} \right) Y \nonumber\\
\left(-3\Box\Psi-\Box\Phi+\f{2}\partial_\mu\partial_\nu h^{(\mu\nu)}+\Box b\right)\nabla_n Y &=&0\nonumber\\
\left(-8\Psi\right) Y_{(mn)} &=&0 \label{eqn:tensor_psi_phi} \\
\left[3\Box\Psi+2\triangle\Psi+\Box\Phi-2B^2\Phi-\f{2}\partial_\mu\partial_\nu
  h^{(\mu\nu)}+\triangle b\right]Y &=&0\nonumber
\eea
Similarly the extra-dimensional component of the equation of motion
for $b^N$ yields
\be\label{equ:eomb}
\left[B ^2\left(-4\Psi+2\Phi\right)-\left(\Box+\triangle\right)b\right]Y=0
\ee
For $\ell \ge 2$ the tensor equation (\ref{eqn:tensor_psi_phi}) sets
to zero the field $\Psi$ that couples to the brane
source.
This confirms our expectation of
zero potential from the exchange of all modes with $\ell \ge 2$.Note that this not the case in arbitrary
  codimension, see \cite{Martin:2004wp} for details.

For $\ell = 1$, however, there is no tensor spherical harmonic
($Y_{(mn);\;\ell=1}  = 0$).  Instead we solve the remaining
equations in the static limit to obtain
\be
\Psi = \frac{1}{10 B^2} \sum_{i=1,\;2} f^4_i \ \f{r^2} Y_{\ell=1,m}(\theta_i, \phi_i)
\ee
Replacing the requisite factors of the 6D fundamental scale
$M$, the potential between two branes of equal tension $f^4$
separated by an angle $\theta$ on the sphere is identical to
that given in Eq.\ (\ref{equ:leq1pot}), confirming
the result obtained by integrating out the $\ell=1$ modes at
tree level.


\section{Non-Perturbative Sphere}\label{app:non-perturbative}
The full metric in the presence of finite-tension branes is 4D de Sitter space with Hubble parameter $H$
orthogonal to a compact 2-sphere of radius $r$:
\[
ds^2 = dt^2 - e^{2Ht} \, \vec x \cdot \vec x - r^2 \left( d\theta + \sin^2\theta \ d\phi \right)
\]
Einstein's equation now contains a contribution due to the
energy-momentum tensor of the branes, $T_{MN}$
\be
R_{MN}- \frac{1}{2}g_{MN}R = \Lambda g_{MN} + \frac{1}{4}g_{MN}F_{OP}F^{OP}-F_{MO}F_N^{\;O} + T_{MN}
\ee
Taking the trace of the 4-dimensional and extra-dimensional
components independently yields
\bea
M^4 \left( 3 H^2 + \f{r^2} \right) &=& \Lambda + \frac{B^2}{2}
                                                        + \sum_i f_i^4 \delta(\vec y) \\
M^4  \left( 6 H^2           \right) &=& \Lambda - \frac{B^2}{2}
\eea
Integrating the first equation over the extra dimensions gives
\be
A \left( 3 M^4 H^2 + M^4 \f{r^2} -  \Lambda - \frac{B^2}{2} \right)
= \sum_i f_i^4 \;\;\;\;,
\ee
relating the sum of the brane tensions to other properties of the
space.  In the absence of branes, by tuning the magnetic field against
the cosmological constant as in Eq.\ (\ref{equ:tuning}), we can choose to have
a static, non-inflationary space.   Turning on a finite tension for a
pair of antipodal branes, each cutting out a deficit angle $\delta =
f^4$ gives football-shaped extra dimensions \cite{Carroll:2003db}, with area
\be
A = 2\left( 2\pi - \delta \right)r^2
\ee

As asserted in \cite{Garriga:2004tq}, since the space is stabilised by a magnetic
flux, it is this flux that must be conserved even if the size of the
space changes.  Holding the magnetic \emph{field}
constant instead leads to the incorrect conclusion
that the 4D vacuum energy only affects the extra-dimensional geometry,
through the global deficit angle, and not the geometry on the branes
\cite{Carroll:2003db,Navarro:2003vw}.  The leading order change in the
Hubble constant and the size of the space, in the presence of the
branes is
\be
r^2 = r_0^2 \left( 1 + \frac{2\delta}{\pi}  + \mathcal O(\delta^2) \right)
\qquad
H^2 = \frac{1}{4 \pi r_0^2} \left( \delta + \mathcal O(\delta^3) \right)
\ee
where $r_0$ is the radius of the sphere with no branes.

A 4D observer would estimate the 4D vacuum energy to be
$\Lambda_4 = 3 M_4^2 H^2$.  At lowest order this is related to
the sum of tensions
\be
\Lambda_4
= 3 M_4^2 H^2
\approx 3 \delta + \mathcal O(\delta^2)
\approx 3 \sum_i f_i^4\;\;,
\ee
as expected from the effective field theory.

\end{document}